\documentclass[prd, aps, 10pt, twocolumn, nofootinbib, superscriptaddress, preprintnumbers, balancelastpage, longbibliography]{revtex4-2}
\usepackage[english]{babel}

\usepackage[dvipsnames]{xcolor}
\usepackage{amsmath, mathtools, physics, xfrac, graphicx, afterpage, subfigure, rotating, multirow, tabularx, booktabs, fancyhdr, theorem, moreverb, euscript, psfrag, slashed, mathtools, makecell, adjustbox, dcolumn, bm, graphics, hyperref, lettrine, aas_macros, color, tikz, soul}
\usepackage[utf8]{inputenc}
\usetikzlibrary{decorations.pathmorphing, fadings, decorations.text}

\usepackage[normalem]{ulem}

\definecolor{darkblue}{rgb}{0.0,0.0,0.75}
\definecolor{darkred}{rgb}{0.6,0.0,0}
\definecolor{darkgreen}{rgb}{0.0,0.6,0.}
\definecolor{nice}{rgb}{0.8,0, 0.8}

\hypersetup{
     colorlinks = true,
     citecolor  = Green,
     urlcolor   = Green,
     linkcolor  = Green
}

\newcommand{\hp}[0]{H$^+_3\;$}

\input Zallman.fd

\LettrineTextFont{\itshape}
\begin{document}

\preprint{MIT--CTP/5748}
\preprint{SLAC-P-24001}

\title{Search for Dark Matter Induced Airglow in Planetary Atmospheres}

\author{Carlos Blanco}
\thanks{\href{mailto:carlosblanco2718@princeton.edu}{carlosblanco2718@princeton.edu}, \href{https://orcid.org/0000-0001-8971-834X}{0000-0001-8971-834X}}
\affiliation{Princeton University, Department of Physics, Princeton, NJ 08544}
\affiliation{Stockholm University and The Oskar Klein Centre for Cosmoparticle Physics, Alba Nova, 10691 Stockholm, Sweden}

\author{Rebecca~K.~Leane}
\thanks{\href{mailto:rleane@slac.stanford.edu}{rleane@slac.stanford.edu}, \href{https://orcid.org/0000-0002-1287-8780}{0000-0002-1287-8780}}
\affiliation{Particle Theory Group, SLAC National Accelerator Laboratory, Stanford, CA 94305, USA}
\affiliation{Kavli Institute for Particle Astrophysics and Cosmology, Stanford University, Stanford, CA 94305, USA}

\author{Marianne Moore}
\thanks{\href{mailto:mamoore@mit.edu}{mamoore@mit.edu}, \href{https://orcid.org/0000-0002-6424-0594}{0000-0002-6424-0594}}
\affiliation{Center for Theoretical Physics, Massachusetts Institute of Technology, Cambridge, MA 02139, USA}

\author{Joshua Tong}
\thanks{\href{mailto:joshtong@stanford.edu}{joshtong@stanford.edu}, \href{https://orcid.org/0000-0002-6818-055X}{0000-0002-6818-055X}}
\affiliation{Particle Theory Group, SLAC National Accelerator Laboratory, Stanford, CA 94305, USA}
\affiliation{Kavli Institute for Particle Astrophysics and Cosmology, Stanford University, Stanford, CA 94305, USA}

\date{\today}

\begin{abstract}

We point out that dark matter can illuminate planetary skies via ultraviolet airglow. Dark matter annihilation products can excite molecular hydrogen, which then deexcites to produce ultraviolet emission in the Lyman and Werner bands. We search for this new effect by analyzing nightside ultraviolet radiation data from Voyager and New Horizons flybys of Neptune, Uranus, Saturn, and Jupiter. {We set new constraints on the dark matter-nucleon scattering cross section for DM above the electron mass; these extend down to $10^{-40}~$cm$^2$ for DM masses around 1 GeV when most of the energy is deposited in the atmosphere}. We highlight that future ultraviolet airglow measurements of Solar System planets or other worlds provide a new dark matter discovery avenue.

\end{abstract}

\maketitle

\lettrine{T}{he night sky} is never truly dark. Even when submerged in complete darkness, the atmosphere casts a faint glow onto the Earth. This phenomenon was observed more than two thousand years ago, when Aristotle ascribed it to \textit{``condensation of the upper air into an inflammable condition"}~\cite{Aristotle,1962JATP...24.1108S}. A few hundred years later, Pliny the Elder also pondered the origin of this effect, calling it \textit{``night suns"}~\cite{Pliny,1962JATP...24.1108S}. Today, we know this effect as \textit{airglow}. At night on Earth, visible airglow dominantly arises due to recombination of molecules which were previously broken by solar radiation, and contributes more light than starlight to the total luminosity of the night sky. 

Airglow is a generic effect in planetary atmospheres. Unlike aurorae, which are correlated with high solar activity and are concentrated near the magnetic poles, airglow occurs globally and is present both day and night, though it is most noticeable at night. While airglow was first observed in the optical, it shines across a wide wavelength range from the infrared (IR) to ultraviolet (UV), occurring due to a variety of photochemical processes including photoionization, photodissociation, chemiluminescence, or molecular excitation by cosmic rays~\cite{2002GMS...130...77S}. Airglow measurements therefore provide insights to the chemical composition and dynamics of planetary atmospheres, as well as space weather.

\begin{figure}[t]
    \centering
    \scalebox{0.86}{\begin{tikzpicture}
        
    \def\rj{2.5} 

    \path[use as bounding box] (-{1.3*\rj},{-1.2*\rj}) rectangle ({1.3*\rj},{1.3*\rj});
    
    \shade [inner color=green, outer color=white, even odd rule] (0,0) circle (\rj+0.9) circle (\rj);

    \fill[color=magenta, path fading=north] (0, -\rj) ellipse (\rj*0.3 and \rj*0.1);
    \fill[color=magenta, path fading=south] (0, \rj) ellipse (\rj*0.3 and \rj*0.1);

     \begin{scope}
       \fill[blue] (0,0) circle (\rj);
       \clip (0,0) circle (\rj);
        \shade[outer color=blue, inner color=cyan] (-0.8,0.95) circle (4);
    \end{scope}

    \node[label=above:{$\chi\chi$}] (A) at (0,0) [circle, fill, inner sep=1.5pt] {};
    \coordinate (B) at ({0.85*\rj*cos(-10)},{0.85*\rj*sin(-10)});
    \draw[dashed] (A) -- node[above] {$\phi$} (B);
    \foreach \i in {-40,-10,20}
        \draw[-latex] (B) -- ++ ({0.7*cos(\i)},{0.7*sin(\i)});

    \node[label=above:{$\chi\chi$}] (C) at ({1.1*\rj*cos(55)},{1.1*\rj*sin(55)}) [circle, fill, inner sep=1.5pt] {};
    \coordinate (D) at ({1.1*\rj*cos(45)},{1.1*\rj*sin(45)});
    \draw[dashed] (C) -- (D);
    \foreach \j in {-70,-20}
        \draw[-latex] (D) -- ++ ({0.4*cos(\j)},{0.4*sin(\j)});
    

    \draw [decorate,decoration={raise=-2ex,text along path,text align=center,text={|\sc\large|Dark Matter Airglow}}] ({-1.05*\rj},-\rj) to [bend left=40]  ({-1.05*\rj},\rj);

    \draw [decorate,decoration={raise=-2ex,text along path,text align=center,text={|\sc|Aurora}}] ({-\rj},0.9*\rj) to [bend left=40]  ({\rj},0.9*\rj);

    \draw [decorate,decoration={raise=-2ex,text along path,text align=center,text={|\sc|Aurora}}] ({-\rj},-0.75*\rj) to [bend right=40]  ({\rj},-0.75*\rj);
        
\end{tikzpicture}}
    \caption{A planet with aurorae (at the poles, magenta), and a dark matter induced ultraviolet airglow (isotropic, green).}
  \label{fig:planet}
\end{figure}
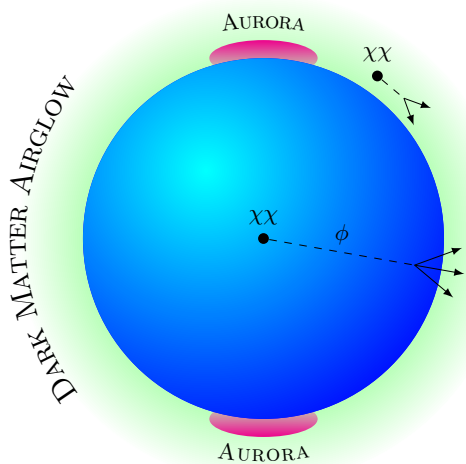

Motivated by the above, the airglow of planets in our Solar System has been measured in exquisite detail by multiple space probes during flybys. In 1977, two NASA spacecraft, Voyager 1 and Voyager 2, launched exploiting a rare planetary alignment occurring once every 175 years, allowing flybys to all the outer planets with gravitational assists. Both probes flew by gas giants Jupiter and Saturn, while Voyager 2 remains today the only probe to have flown by ice giants Uranus and Neptune. Other space probes detected airglow in the giant planets, including Galileo, New Horizons, and Cassini~\cite{1992PhDT........12H, 2004jpsm.book..185Y}.

In this \textit{Letter}, we point out for the first time that dark matter (DM) annihilation can induce UV airglow in planetary atmospheres. This occurs if DM annihilation products excite molecular hydrogen, which then deexcites to produce UV emission in the Lyman and Werner bands. This would produce an isotropic UV airglow across planets, as schematically shown in Fig.~\ref{fig:planet}. We analyze Voyager and New Horizons UV spectral data on the gas and ice giants to search for this new signature, allowing a new test of DM interactions with the Standard Model. The benefit of using flyby data is that night airglow measurements, which have low background and cannot be obtained from Earth, are available. We will show that this provides leading sensitivity to the particle nature of DM and a promising discovery method.

We focus on the Solar System gas and ice giants since (i) their large masses and radii lead to accumulation of a large DM population, and thus their DM annihilation rate is large, (ii) they are far from the Sun, resulting in low solar-induced airglow background, (iii) they have been visited by space probes providing night airglow measurements, and (iv) they are relatively cold, so their thermal UV emissions do not overshadow the DM-annihilation signal. We consider all four giant planets, as they vary in their overall number of captured DM particles and radial DM distributions due to differences in their density, temperature, mass, and radius. Additionally, space probe measurements differ in precision, depending on factors like distance, spatial and spectral resolution, and solar activity at the time of flyby. Considering these differences, we expect complementarity in studying DM annihilation across these objects, both in the DM model space and data interpretation. In a companion paper~\cite{Blanco:2025wpo}, we investigate the application of our signal to different model classes and complementary probes. 

Our paper is organized as follows. We first discuss
how UV airglow arises and is detected, before discussing existing Voyager and New Horizons observational data and our analysis. We then detail how DM-induced airglow can be calculated in Neptune, Uranus, Saturn, and Jupiter, and determine the maximum DM airglow consistent with Voyager and New Horizons UV data. Next we explore the relevant DM-nucleon scattering parameter space, setting new constraints based on the requirement that DM-induced airglow would exceed flyby measurements. We conclude with future opportunities to discover DM using planetary airglow. \\

\noindent\textbf{\textit{Overview of Hydrogen Airglow--}} 
The upper atmospheres of the giant planets are dominantly composed of molecular hydrogen, helium, and hydrocarbons (mainly methane). Precipitating magnetospheric electrons bombard the atmosphere, exciting these species and causing them to emit in the UV band ($700-1800$~\r{A}), observable as aurorae and contributing to non-polar non-auroral emissions (airglow). About half of the observed airglow of the gas giants is caused by hydrogen fluorescence and secondary photoelectrons following the absorption of solar light~\cite{2010Icar..210..270G, 1996ApJ...462..502L}. As the electronically excited gas decays to its ground state, the UV emissions are proportional to the intensity of the precipitating electrons as well as the intensity of the solar extreme UV radiation~\cite{2010Icar..210..270G, 1996ApJ...462..502L, 2012JGRA..117.7316G}. 

The UV spectrum of atmospheric hydrogen is dominated by atomic H Lyman-$\alpha$ (HLy$\alpha$) lines and vibronic H$_2$ Lyman and Werner bands. The vibronic bands correspond to UV emissions from lowest-lying excited electronic states in molecular hydrogen~\cite{dabrowski1984lyman}. These excited states radiatively deexcite to the totally symmetric ground state with radiative lifetimes that are representative of efficient electronic transitions, ${\tau \sim \mathcal{O}(\text{ns})}$~\cite{salumbides2015h2}. Deexcitation to the vibrational continuum can occur, dissociating the molecule into two neutral atoms. This occurs with an efficiency of about 10\% and generates a continuum emission spectrum~\cite{abgrall1997emission,2004Icar..171..336G,2013JMoSp.291..108G}.
The continuous presence of H$_2$ Lyman and Werner lines indicates the steady-state presence of electronically excited molecular hydrogen due to the high efficiency of its radiative deexcitation. For more details on the Lyman and Werner bands, see the Supplemental Material. 

While the H Lyman-$\alpha$ line is the most salient spectral feature, {it} makes up less than 10\% of the total unabsorbed H$_2$ emission in the auroras of Jupiter and Saturn~\cite{2012JGRA..117.7316G}, and less than 20\% of the total UV spectrum of their airglow~\cite{2010Icar..210..270G, 1996ApJ...462..502L}. We focus on the H$_2$ Lyman and Werner bands over H Lyman-$\alpha$ since the H Lyman-$\alpha$ signal often includes substantial contributions from beyond the atmosphere, $e.g.$ the planet's rings and magnetosphere~\cite{1981Sci...212..206B}. The heliospheric H Lyman-$\alpha$ background is also of comparable size or larger than the airglow, complicating the isolation of the planetary signal~\cite{1992PhDT........12H, 2018GeoRL..45.8022G}.

Deeper in these atmospheres, hydrocarbons such as methane, ethane, and acetylene become abundant enough to absorb shorter-wavelength UV emissions from hydrogen. This leads to a slightly modified spectra, mainly at wavelengths below $1350$~\AA~\cite{2007Sci...318..229G, 2012JGRA..117.7316G}. In order to quantify the spectral distortion due to the attenuation of UV emission, a color ratio parametrizes the hydrocarbon absorption at lower wavelengths compared to the non-attenuated {brightness} at longer wavelengths, and is therefore a proxy for the altitude of the auroral emission. This is defined as the relative intensities of the measured spectra in two spectral windows, specifically,
\begin{equation}
    \text{Color Ratio} = \frac{\text{I}(1550-1620~\text{\r{A}}) }{\text{I}(1230-1300~\text{\r{A}})}\ .
\end{equation}
This effect is more pronounced in Jupiter than in Saturn~\cite{2012JGRA..117.7316G}. An unattenuated spectrum yields a color ratio of 1.1 and 1.5 for auroras and airglow, respectively. While generally variable between 2 and 10, a typical value for the CR in Jupiter's aurora 2.5~\cite{2012JGRA..117.7316G, 2004Icar..171..336G,2013JMoSp.291..108G}. 

Solar radiation or cosmic rays can also \textit{ionize} atmospheric hydrogen, leading to secondary IR-UV emissions that can further probe atmospheric energy deposition. Ionized H$_2$ leads to the formation of H$_3^+$, which emits strongly in the IR between about 3~$\mu$m and 5~$\mu$m, making it an efficient thermal radiator at temperatures between 700~K and 1000~K~\cite{RevModPhys.92.035003,miller2006driver}. This is the main thermostatic mechanism in Jupiter, where H$_3^+$ IR emissions are a measure of ionizing power deposition in the upper atmosphere~\cite{Blanco:2023qgi}. Although a powerful probe of DM-induced ionization on Jupiter~\cite{Blanco:2023qgi}, it is less applicable to the other giant planets' atmospheres due to differing temperature ranges and the uncertain survivability of H$_3^+$, as well as the validity of local thermal and chemical equilibrium. In contrast, far-UV hydrogen emissions are essentially temperature independent, providing an indirect measurement of the power injected by radiation in the atmosphere of \textit{all} giant planets. \\

\noindent\textbf{\textit{Voyager and New Horizons UV Airglow Data--}}
We use Voyager 1, 2, and New Horizons mission data, which provide the best probes of UV night airglow (also known as nightglow) in the giant planets. To minimize the background in our analysis, we focus on the nightglow away from the polar aurorae, dawn, and dusk, where scattered solar irradiation can influence measurements. We will first set a limit on additional H$_2$ emission, $i.e.$ determine the brightest nightglow consistent with data, and in the next section use it to constrain DM.

{For Jupiter, we use the processed and background-subtracted data of the Jovian nightglow from four scans of New Horizons' UV imaging spectrograph~\cite{2007Sci...318..229G, 2011epsc.conf.1474B}. In order to translate the UV spectrum of Jupiter measured by New Horizons (Figure~\ref{fig:NH_spec}) into an emission brightness in the spectral window [924 $-$ 1072]~\r{A}, we account for the atmospheric absorption of the intrinsic UV brightness, and then integrate over the spectral ROI.} 

{Figure~\ref{fig:NH_spec} shows the de-absorbed spectrum that we adopt as our model of the UV-airglow emission for Jupiter in our spectral ROI (details in Supplemental Material). We find a Jovian nightglow H$_2$ brightness of 3.1~rayleighs~(R) (one rayleigh is the apparent emission rate of $10^{10}$ photons/$(4\pi)$/m$^2$/s/sr~\cite{doi:https://doi.org/10.1002/9781118668047.app2}), with a brightness of 5R saturating the range of the measured data.}  



In Fig.~\ref{fig:NH_spec}, the spectral features in the data are likely a combination of noise and detected spectral lines. On the other hand, the synthetic spectra have not been convolved with the instrument response and show substructure coming from the modeled deexcitation of many narrow rovibronic states~\cite{2004Icar..171..336G,2013JMoSp.291..108G}. Since we consider the integrated brightness over a spectral window that is large compared to the spectral features, we find that our results are insensitive to this substructure. Note that this is not a likelihood analysis and the data range should not be interpreted as a confidence range; however given we set our constraint on the requirement that the signal fully saturates the data, this is a conservative approach.

\begin{figure}[t!]
    \centering
    \vspace{-2mm}
    \includegraphics[width=0.9\columnwidth]{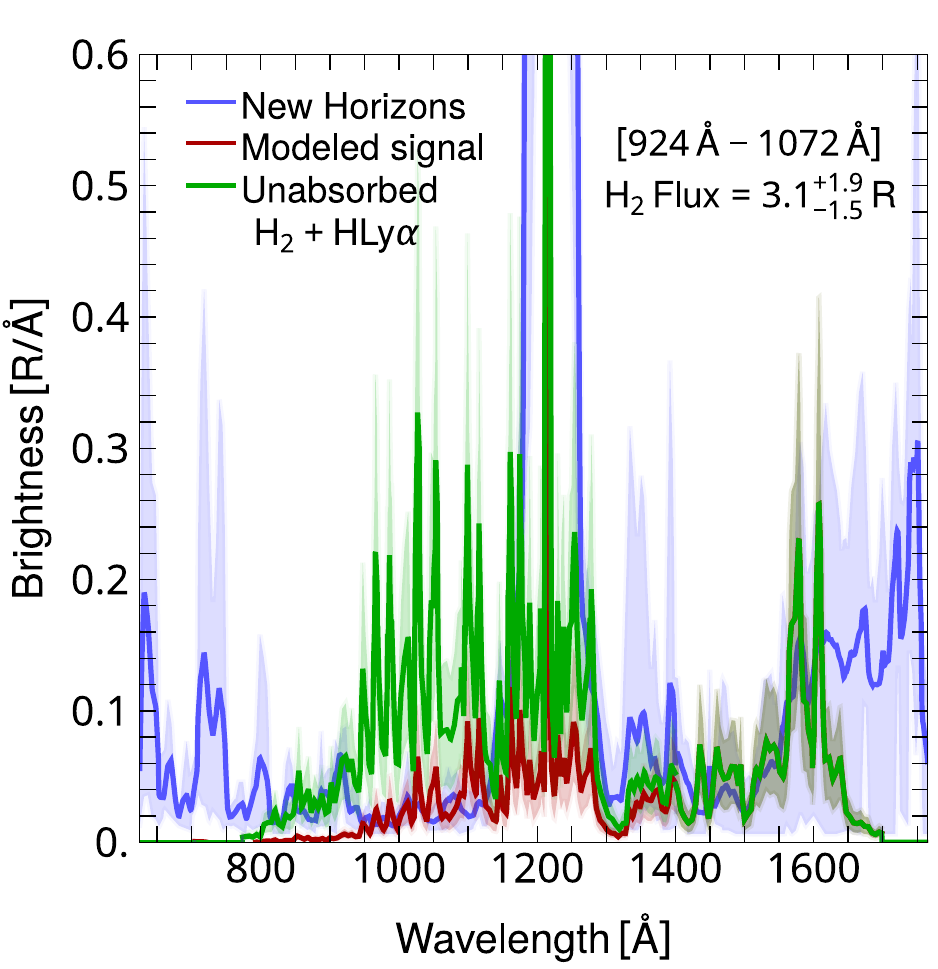}
    \vspace{-10pt}
    \caption{New Horizons flyby scans of Jovian UV nightglow (range is shaded blue, mean is solid blue), overlayed with a synthetic H$_2$ emission spectrum without any atmospheric hydrocarbon absorption (green), as well as the true expected spectrum corrected for attenuation from absorption, shown at the maximum value consistent with the data (red).}
    \label{fig:NH_spec}
\end{figure}

\begin{table}[]
    \centering
    \caption{Measurements of the H$_2$ Lyman and Werner {brightness} from the nightside equator region of the Solar System's giant planets, and corresponding input power into the planet's upper atmosphere. Uranus and Neptune's {brightness} includes the Rydberg bands from H$_2$ and may originate from the dayside\footnote{{No uncertainty is reported in Uranus' original measurement. An uncertainty comparable to that of Neptune ($\sim 16\%$) would not affect our conclusions.}}.}
    \begin{tabular}{p{1.0cm} r r r p{0.6cm}}
        \toprule 
        Planet & {Brightness} [R] & Power [$\mu$W/m$^2$] & Space Probe & Ref. \\ \midrule
        Jupiter & $3.1^{+1.9}_{-1.5}$ & $0.31^{+0.19}_{-0.15}$ & New Horizons & \cite{2011epsc.conf.1474B} \\
        Saturn & $<10$ & $<1$ & Voyager 1 & \cite{1981Sci...212..206B} \\
        Uranus & $46$ & $4.6$ & Voyager 2 & \cite{1989Sci...246.1459B} \\
        Neptune & $19\pm3$ & $1.9\pm0.3$ & Voyager 2 & \cite{1989Sci...246.1459B} \\ \bottomrule 
    \end{tabular}
    \label{tab:H2_Lyman_planets}
\end{table}

For the other giant planets, UV nightglow measurements are already readily available, and do not require additional analysis. For Saturn, Voyager 1 detected H$_2$ Lyman and Werner emission in its disk and auroral regions, but not on its nightside~\cite{1979Sci...206..962S, 1981Sci...212..206B, 1982Sci...215..548S}. H Lyman-$\alpha$ emission was detected on its nightside, but potentially contains contamination from Saturn's rings and magnetosphere. We adopt the quoted upper limit of 10~R for H$_2$ night airglow emission from Saturn~\cite{1979Sci...206..962S}. Uranus and Neptune were only visited by Voyager 2, which was optimized to study Jupiter and Saturn, resulting in fewer and less detailed observations of the ice giants~\cite{1986Sci...233...74B, 1987RvGeo..25..251B, 1989Sci...246.1459B}.

Table~\ref{tab:H2_Lyman_planets} summarizes measurements of H$_2$ Lyman and Werner {brightness} from the nightside equatorial regions of the Solar System's giant planets, alongside the corresponding input power into their upper atmosphere. The collected data is H$_2$ emission brightness. Multiple studies have explored the relationship between the emission brightness observed in the H and H$_2$ bands, and the power injected by precipitating electrons in a giant planet's atmosphere~\cite{1982JGR....87.4525G, 1983JGR....88.6143W, 2012JGRA..117.7316G}. A widely accepted correspondence in the planetary UV literature is that 10~R of H$_2$ emission in the Lyman and Werner bands correlates with 1~$\mu$W/m$^2$ of input power for electron precipitation. This relationship allows us to estimate upper limits on power injected into the planet's upper atmosphere from electron precipitation, and therefore to set upper limits on injection of high-energy electrons from an additional source, such as DM.\\

\noindent\textbf{\textit{Dark Matter Induced UV Airglow--}}
As the giant planets traverse the galaxy's DM halo, they can trap and accumulate DM particles. The DM can then annihilate into electrons, exciting molecular hydrogen which deexcites with a spectral shape as shown in Fig.~\ref{fig:NH_spec}. To calculate the normalization of this spectrum for a range of DM parameters, we take the DM mass annihilation rate~\cite{Leane:2020wob},
\begin{align}\label{eq:dm_ann}
    \Gamma_\text{ann} &= \pi R^2 \sqrt{\frac{8}{3\pi}} \rho_\chi v_\chi \left( 1 + \frac{3\, v_\text{esc}^2}{2\,v_\chi^2} \right)\times f_{\rm cap}\ .
\end{align}
Here ${\rho_\chi = 0.4}$~GeV/cm$^3$ is the local DM density, ${v_\chi = 270}$~km/s is the local DM velocity, $R$ is the planet radius, and $v_\text{esc}$ is the planet's escape velocity. We neglect motion relative to the halo, which is a minor correction. $f_{\rm cap}$ is the fraction of DM particles captured by the planet, which scales with the scattering cross section. We calculate the capture rate as a function of DM mass and scattering cross sections using the \texttt{Asteria} package~\cite{asteria, Leane:2023woh}, and use the planetary modeling inputs as detailed in the Supplemental Material. Note that Eq.~\eqref{eq:dm_ann} assumes capture-annihilation equilibrium, which will not apply for all DM models especially at low cross sections; we investigate this in our companion paper~\cite{Blanco:2025wpo}.

To produce detectable airglow, the DM annihilation products must be deposited into the atmosphere. {Even if DM annihilates directly into electrons, not all of the injected energy contributes to the excitation of atmospheric hydrogen. We account for this by introducing an additional deposition efficiency, $f_\text{dep}(m_\chi)$, evaluated at the characteristic electron energy set by the DM mass. We obtain the collisional and total stopping powers of electrons propagating through hydrogen as functions of electron energy using the ESTAR database from NIST~\cite{ESTAR}. The observable DM-induced injected airglow power is therefore}
\begin{align}\label{eq:dmpower}
   P_\text{atm}^\text{DM} &= \frac{\Gamma_\text{ann} \times f_\text{atm} {\times f_\text{dep}}}{4 \pi R^2} \ ,
\end{align}
where $f_\text{atm}$ is the fraction of annihilation events injecting energy into the atmosphere, which can take a range of values depending on the DM model. If DM annihilates in the core, and the annihilation products are
boosted or produced through decay of a long-lived mediator, ${f_\text{atm}\sim1}$ is achieved across a wide range of parameters~\cite{Batell:2009zp,Pospelov:2007mp,Pospelov:2008jd,Rothstein:2009pm,Chen:2009ab,Schuster:2009au,Schuster:2009fc,Bell_2011,Feng:2015hja,Feng:2016ijc,Allahverdi:2016fvl,Leane:2017vag,Arina:2017sng,Albert:2018jwh, Albert:2018vcq,Nisa:2019mpb,Niblaeus:2019gjk,Cuoco:2019mlb,Serini:2020yhb,Mazziotta:2020foa, Leane:2021ihh, Bell:2021pyy,Leane:2021tjj,Li:2022wix,Linden:2024uph, Leane:2024bvh,Acevedo:2023xnu}. Alternatively, a large population of DM particles can sit thermalized in the atmosphere~\cite{Leane:2022hkk}, such that heavy mediators can also provide new sensitivities~\cite{Blanco:2025wpo}.  We assume ${f_\text{atm}\sim1}$ and investigate other choices in the companion paper~\cite{Blanco:2025wpo}. {Because the injected power scales linearly with $f_\text{atm}$, limits relax by a factor ${1 / f_\text{atm} }$ when ${f_\text{atm}<1}$.}

We set a limit on DM-induced airglow when the DM-power of Eq.~\eqref{eq:dmpower} exceeds the power derived from Voyager and New Horizons data on the H$_2$ Lyman and Werner bands of Tab.~\ref{tab:H2_Lyman_planets}. In comparing with Tab.~\ref{tab:H2_Lyman_planets}, our assumption is that the relationship between emission brightness and power input is the same for both DM annihilation and astrophysical electron precipitation. For DM annihilation into electrons, this is accurate. For DM annihilation into final states that are not electrons (other than neutrinos), the DM limit is only reduced by a factor of a few, as other final states produce energetic electrons through $e.g.$ bremsstrahlung or decay~\cite{Cirelli:2010xx}.\\

\noindent\textbf{\textit{New Dark Matter Sensitivities--}} 
Figure~\ref{fig:H2_LW} shows our new constraints on the DM mass as a function of spin-dependent scattering cross section for the gas and ice giants, using Voyager and New Horizons UV night airglow data. These constraints are presented for pure proton couplings, with additional cases detailed in the Supplemental Material. The limits exhibit enhancements when the DM mass aligns with the mass of planetary atoms -- around hydrogen for Jupiter and Saturn, and near hydrogen and nitrogen for Uranus and Neptune.

\begin{figure}[t]
    \centering
    \includegraphics[width=\columnwidth]{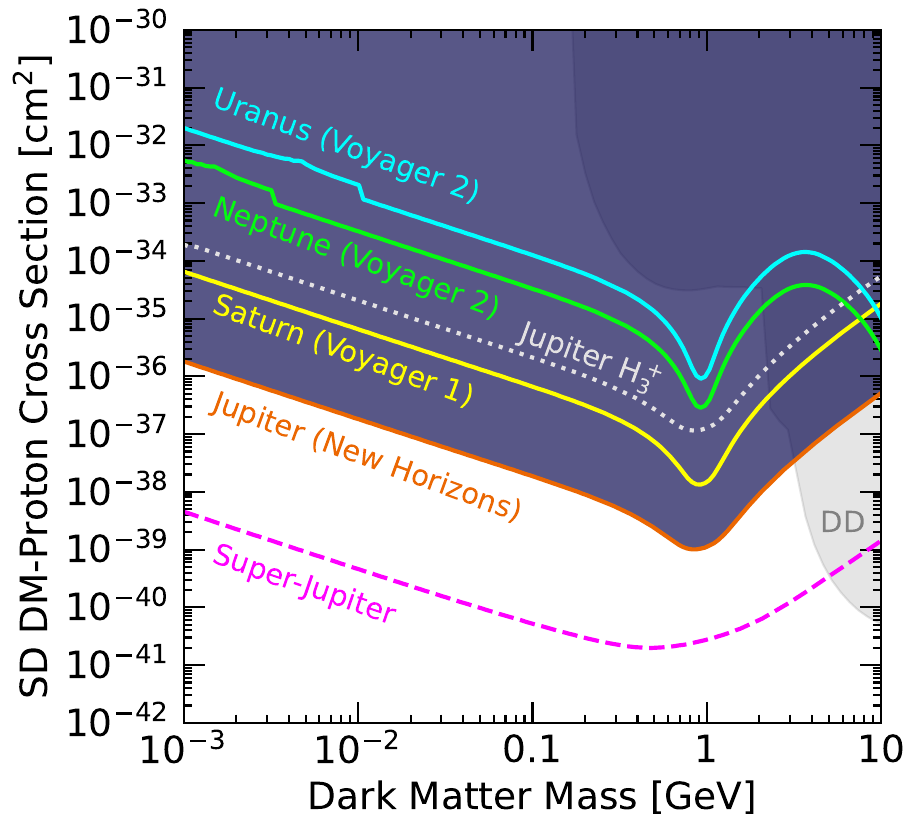}
    \caption{New constraints on the DM mass vs scattering cross section using Voyager and New Horizons data on Solar System gas and ice giants as labeled{, for $f_\text{atm} \sim1$}. We also show projected sensitivity for a nearby Super-Jupiter. Complementary constraints from \hp production in Jupiter and direct detection (DD) experiments are shown as labeled.}
    \label{fig:H2_LW}
\end{figure}

Our results from the four giant planets are highly complementary. In particular, because the radii and interior profiles (density and temperature) of these objects differ, their results apply to different particle model classes. For example, each planet is optimized for different dark-visible sector mediator masses and kinematic boosts, as the DM annihilation products may be deposited at different radii within the objects. In addition, their differing density and temperature profiles lead to different internal distributions of DM, such that complementary DM masses and cross sections can be probed in the case of heavy mediator models. We discuss in detail the interplay of model classes probed by our new search in a companion paper~\cite{Blanco:2025wpo}; in Fig.~\ref{fig:H2_LW} we have focused on the scenario where all DM annihilation products are deposited in the given planet's atmosphere.

In Fig.~\ref{fig:H2_LW}, we show projected sensitivity to DM airglow for a benchmark local Super-Jupiter, with a mass ten times Jupiter's, and the same radius. The enhanced sensitivity largely arises as our benchmark is more massive than Solar System planets, leading to a larger capture rate. We assume the planet is either on a wide orbit or is free-floating, avoiding stellar irradiation. We therefore take the dominant background to be cosmic-ray electron irradiation, which locally is $1.25~\mu$W/m$^2$~\cite{2018A&A...614A.111P}. We assume auroral and airglow signals cannot be distinguished, requiring the DM signal to exceed the background everywhere. While even brighter DM airglow would be produced for an inner Galaxy planet where the ambient DM density is larger, such distant UV signals are difficult to observe due to interstellar medium absorption; {this benchmark planet therefore likely must be local for detection.}

Figure~\ref{fig:H2_LW} also presents our recast bound on DM from \hp production in Jupiter of $57~\mu$W/m$^2$~\cite{Blanco:2023qgi}, using improved modeling of Jupiter's interior {and deposition efficiency $f_\text{dep}$}. Our UV airglow results surpass those from \hp by more than two orders of magnitude. In addition, as discussed earlier, UV airglow applies more broadly than \hp production to planets beyond Jupiter. Given the complementary interpretation of constraints from the giant planets as discussed above and shown in a companion paper~\cite{Blanco:2025wpo}, our new sensitivities are not only stronger but also apply to a wider range of DM models than the Jovian \hp constraints.

While there are differing model assumptions, in Fig.~\ref{fig:H2_LW}, we also include complementary constraints on spin-dependent proton-DM scattering from direct detection experiments~\cite{Cirelli:2024ssz}, including {PICO-60}~\cite{PICO:2019vsc}, {NEWS-G}~\cite{NEWS-G:2024jms}, {CRESST-III}~\cite{CRESST:2022dtl}, {PICASSO}~\cite{PICASSO:2012ngj}, and {LZ}~\cite{LZ:2024zvo}. {By contrast, comparison with indirect detection constraints are typically strongly model-dependent and are therefore not shown.} Similarly, DM evaporation is model dependent~\cite{Acevedo:2023owd}, and is not included in Fig.~\ref{fig:H2_LW}. We apply our findings to some model classes in a companion paper~\cite{Blanco:2025wpo}.

There are complementary DM constraints from Jovian gamma rays~\cite{Batell:2009zp,Leane:2021tjj, Linden:2024uph, Leane:2024bvh}. These are not directly comparable, as they rely on detecting the gamma-ray spectral energy distribution, which is model dependent~\cite{Elor:2015bho,Leane:2024bvh}. Our bounds instead only rely on the total power injected, and therefore are more generic. Our bounds are also complementary to those set from
electrons detected in Jupiter’s radiation belt, as
our results do not depend on the details of the time-varying magnetic field around Jupiter~\cite{Li:2022wix}. There are also existing constraints using celestial bodies in the Galactic Center~\cite{Leane:2021ihh,Acevedo:2023xnu,John:2023knt,Linden:2024uph,Acevedo:2024ttq,Leane:2024bvh}; however, in addition to varying with the final state energy distribution in some cases, there are large systematics in the DM density. Our constraints instead rely on the local DM density.\\

\noindent\textbf{\textit{Summary and Future Prospects--}} We pointed out and executed a new search for DM using UV airglow in planets. We used existing flyby observational UV airglow data on the gas and ice giants on our Solar System: Voyager 1 data on Saturn, Voyager 2 data on Uranus and Neptune, and New Horizons data on Jupiter. The giant planets are ideal as being furthest from the Sun they have the least solar irradiation, and in addition are superior DM captors, such that overall they optimize our signal over background. Our new night UV airglow constraints are stronger than other existing constraints on the DM-nucleon scattering cross section.

Our new DM signature offers a promising avenue to detect or constrain DM moving forward. We highlighted future sensitivities assuming UV spectral detection of a nearby Super-Jupiter, which would improve the reach further. Future data from potential upcoming UV telescopes and instruments such as the LUMOS (LUVOIR Ultraviolet Multi-Object Spectrograph) instrument in the LUVOIR-NASA mission concept, or the Habitable Exoplanet Imaging Mission (HabEx), may provide fruitful insights to DM-induced UV airglow. Studying archival Galaxy Evolution Explorer (GALEX) or Hubble Space Telescope (HST) data may also be useful to reveal DM airglow in planetary atmospheres.

Within the Solar System, upcoming opportunities can increase the sensitivity to DM through UV airglow. Currently, the Juno probe is orbiting Jupiter. However, Juno has a low duty cycle because it spins, which makes detecting nightglow difficult, as it needs to be differentiated from bright aurorae. Another spacecraft, JUICE (Jupiter Icy Moons Explorer)~\cite{2023SSRv..219...53F}, just launched in 2023 and is on its way to Jupiter with an onboard UV spectrometer. JUICE observations will be at larger distances than Juno (about 10 Jupiter radii away), and therefore will not have superior spatial resolution over Juno, but JUICE may offer the best opportunity for deep nightglow spectra {due to its pointed observing strategy and longer integration times. While such observations could achieve sensitivities comparable to or better than those quoted in Table~\ref{tab:H2_Lyman_planets}, a quantitative forecast depends on mission-specific parameters such as observing geometry, background systematics, and solar activity, and requires dedicated simulations beyond the scope of this work.} For the ice giants of the Solar System, the only spatially resolved UV data from Uranus comes from the Voyager 2 encounter in 1986. The proposed Uranus Orbiter and Probe may gather more data about Uranus' atmospherical processes in the upcoming decades~\cite{2022PSJ.....3...58C}. {Overall, a consistent signal across all four giant planets would provide a powerful crosscheck, especially given the varying precision of space probe observations due to differences in distance, resolution, and solar activity.}\\

\noindent\textbf{\textit{Acknowledgments--}} 
We thank Randy Gladstone, Wayne Pryor, and Jacques Gustin for helpful discussions. C.B. is supported in part by NASA through the NASA Hubble Fellowship Program grant HST-HF2-51451.001-A awarded by the Space Telescope Science Institute, operated by the Association of Universities for Research in Astronomy, Inc., for NASA, under contract NAS5-26555. R.K.L. and J.T. are supported by the U.S. Department of Energy under Contract DE-AC02-76SF00515. M.M. is supported by subMIT at MIT Physics, the Fonds de Recherche du Québec – Nature et Technologies (FRQNT) doctoral research scholarship (Grant No.~305494), and the Natural Sciences and Engineering Research Council (NSERC) Canada Postgraduate Scholarship - Doctoral (Grant No.~577851). M.M is also supported by the U.S. Department of Energy under contract DE-SC0012567.

\clearpage
\newpage
\maketitle
\onecolumngrid
\begin{center}
\textbf{\large Search for Dark Matter Induced Airglow in Planetary Atmospheres}

\vspace{0.05in}
{ \it \large Supplemental Material}\\ 
\vspace{0.05in}
{Carlos Blanco, Rebecca K. Leane, Marianne Moore, and Joshua Tong}
\end{center}
\onecolumngrid
\setcounter{equation}{0}
\setcounter{figure}{0}
\setcounter{section}{0}
\setcounter{table}{0}
\setcounter{page}{1}
\makeatletter
\renewcommand{\theequation}{S\arabic{equation}}
\renewcommand{\thefigure}{S\arabic{figure}}
\renewcommand{\thetable}{S\arabic{table}}

\tableofcontents

 \section{Overview of the Lyman and Werner Bands}

For our DM-induced UV airglow, the relevant electronic excitation and emission from the H$_2$ Lyman band is given by the process
\begin{eqnarray}
    e^- + \text{H}_2(X^1 \Sigma_g^+) \to e^- + \text{H}_2(B^1 \Sigma_u^+)\\
    \text{H}_2(B^1 \Sigma_u^+) \to \text{H}_2(X^1 \Sigma_g^+) + h\nu\ ,
\end{eqnarray}
and from the H$_2$ Werner band is
\begin{eqnarray}
    e^- + \text{H}_2(X^1 \Sigma_g^+) \to e^- + \text{H}_2(C^1 \Pi_u)\\
    \text{H}_2(C^1 \Pi_u) \to \text{H}_2(X^1 \Sigma_g^+) + h\nu\ .
\end{eqnarray}
For the molecular term symbols shown, the latin letters index the electronic excitations in order of increasing energy, and the greek letters are generalizations of the $s$, $p$, $d$, ... $etc.$ atomic representations of orbital angular momentum quantum numbers. The \textit{1} in the superscripts specifies that the states are spin singlets, while the \textit{(u) g} subscripts specify (un)gerade central inversion symmetry. The + superscript on the $\Sigma$ states specifies the symmetry under reflection through the plane containing the internuclear axis. 

Since the vibronic excited states displace the nuclei far from their equilibrium separation, deexcitations from the $B^1$ and $C^1$ states can occur into the vibrational dissociation continuum of the $X^1 \Sigma_g^+$ ground state, creating two neutral H(1s) atoms as follows,
\begin{eqnarray}
    e^- + \text{H}_2(X^1 \Sigma_g^+) \to e^- + (\text{H}_2(C^1 \Pi_u) \ \text{or}\ \text{H}_2(B^1 \Sigma_u^+))\\
    (\text{H}_2(C^1 \Pi_u) \ \text{or}\ \text{H}_2(B^1 \Sigma_u^+)) \to 2\, \text{H}(1s) + h\nu\ . 
\end{eqnarray}
Note that since the dissociation leaves two atoms with unquantized momenta, the emission spectra is a continuum. 

H$_2$ solar fluorescence is expected to contribute a large fraction of the UV airglow on the day side of the gas giants. This fluorescence signal is caused by the absorption of solar UV photons by H$_2$, followed by rapid radiative deexcitation. In particular, solar emission features from H (mostly Ly $\beta$, $\gamma$, and continuum), oxygen, carbon, and nitrogen coincide with absorption lines in the Lyman and Werner bands of H$_2$ leading to efficient radiative pumping. It should be noted that since some of the solar fluorescence emission happens through absorption by vibronically excited H$_2$, the spectra is dependent on temperature. Since we only consider a signal from charged particle collisions, and since we focus on planetary night sides, we do not include solar fluorescence in our analysis.

\section{Additional Details on UV Measurements of Jupiter and Data Analysis}
\label{app:NW_Jupiter}

Voyager 1 and 2 ultraviolet spectrometers (UVS) have been the most stable far and extreme UV instruments over the course of space exploration history~\cite{2016ApJ...823..161B}. They were first calibrated in the laboratory, and following their launch, were pointed periodically at four bright stars to re-calibrate in flight and identify potential changes in the instrument response~\cite{1992PhDT........12H}, a method which works across the UV spectrum (except for H Lyman-$\alpha$). Voyager 2's instruments retained their performance to within $\pm 10\%$~\cite{1981JGR....86.8259B}. In contrast, during Voyager 1's passage in Jupiter's inner magnetosphere, its UVS deteriorated due to radiation-induced damage, but quickly stabilized and was re-calibrated using Voyager 2's results, as well as routine stellar calibration. New Horizons' UV imaging spectrograph, ALICE, was also calibrated on the ground before launch, including tests for the detector's dark count rate, wavelength calibration, and spectral resolution. The in-flight sensitivity performance was monitored through star observations~\cite{2008SSRv..140..155S}.

{
To model our signal, we adopt a synthetic H$_2$ spectral model representative for Jupiter, as measured in auroral spectra separately by Voyager~1~\cite{2013JMoSp.291..108G}. As discussed in the main text, hydrocarbons absorb UV emission at the lower wavelengths; this is included in the spectrum by taking a color ratio of 2.5~\cite{2012JGRA..117.7316G, 2013JMoSp.291..108G}. Figure~\ref{fig:NH_spec} shows the New Horizons data we use in our Jovian analysis, with the range of measured spectra in blue, and the mean as the solid line. The synthetic H$_2$ spectrum is overlaid to show the maximum H$_2$ signal consistent with the New Horizons data, within the same spectral window used by Voyager 1 to characterize the Jovian H$_2$ emission spectrum, [924 $-$ 1072]~\r{A}~\cite{PDS_catalog}. We show the spectrum both with and without attenuation by hydrocarbon absorption; the red contains the atmospheric absorption and is the expected physical signal shape. For both signal and background, we integrate over the spectral window [924 $-$ 1072]~\r{A} to obtain a total emission brightness. Once our model signal exceeds the background, we set a limit on additional H$_2$ emission. We find that a Jovian nightglow H$_2$ brightness of 3.1~R is most consistent with the data, with a brightness of 5~R saturating the range of the measured data.}

{
We note that both Voyager probes took measurements of the UV spectrum of Jupiter~\cite{1979Sci...204..979B, 1979Sci...206..962S, 1980Icar...43..128M, 1981JGR....86.8259B}. However, these measurements were taken during a Solar maximum, in contrast to the New Horizons measurements at Solar minimum ---giving a smaller background-subtracted apparent brightness.}

 
\section{\label{app:Planet_properties}Properties of the Giant Planets}


Table~\ref{tab:planet_data} outlines the radius, mass, and composition of the gas and ice giants used to obtain the DM capture rate. The planetary radius is defined as the boundary where the internal pressure falls below 1~bar. Regions with smaller radii are considered the planet's interior, while larger radii, where data is available, correspond to the atmosphere.

The atmospheric composition of the giant planets is generally well understood, but their internal elemental abundance is more challenging to constrain. For Jupiter and Saturn, we assume a composition of approximately $75\%$ hydrogen and $25\%$ helium by mass~\cite{2012ApJS2025F}. This assumption is based on the understanding that only the small core regions (${r<0.2~R_J}$, ${r<0.25~R_S}$~\cite{1982AREPS..10..257S}) are likely to contain heavier elements, possibly in the form of silicate rock or ices. Recent models estimate that about ${5-10\%}$ of Jupiter's interior is composed of heavier species, such as carbon, nitrogen, and sulfur~\cite{2010SSRv..152..423F}. While Jupiter's and Saturn's atmosphere may be slightly richer in hydrogen than their interior, with Saturn's hydrogen content potentially reaching up to $89\%$~\cite{1982AREPS..10..257S}, for simplicity we assume a uniform composition throughout the gas and ice giants. We assume that the Super-Jupiter has the same composition as Jupiter and Saturn.

Uranus and Neptune have a significantly different composition. These planets are composed largely of heavy elements, with their interior dominated by ices such as water (H$_2$O), methane (CH$_4$), and ammonia (NH$_3$). However, the exact proportions and phases of these materials remain uncertain~\cite{1987RvGeo..25..251B}. Due to the low temperatures in the outer Solar System, these molecules are likely in condensed phases, forming a dense, icy mantle around a rocky core. Their atmosphere likely consists predominantly of hydrogen and methane~\cite{1982AREPS..10..257S}, with methane contributing to the characteristic blue color of these planets.

Table~\ref{tab:isotopes} details the spin, and expectation value of the proton and neutron spins, for elements relevant for spin-dependent scattering. For elements heavier than hydrogen and helium, we expect their isotopic ratios to be similar across all planets, reflecting their primordial composition. We use the isotopic ratios in the giant planets as reported in the literature~\cite{DEBERGH1995427, 2018P&SS..155...12M}. The values of spin and expectation values are from Refs.~\cite{Bednyakov:2004xq, Hooper:2018bfw}. As the spin information is only relevant for spin-dependent scattering, elements such as ${}^4$He, ${}^{12}$C, and ${}^{16}$O, which have ${J=\ev*{S_{\lbrace p,n \rbrace}} = 0}$ (and so do not contribute to the spin-dependent cross section) are excluded from the table.

\begin{table}[h!]
    \centering
    \caption{List of parameters used for the giant planets: radius $R$, mass $M$, and elemental abundance of the most common elements and isotopes in mass fraction.}
    \begin{tabular}{l r r r r}
         \toprule 
         & Jupiter & Saturn & Uranus & Neptune \\ \midrule
         $R$ ($\times 10^7$ m) & 6.99 & 5.83 & 2.54 & 2.46 \\
         $M$ ($\times 10^{26}$ kg) & 18.98 & 5.68 & 0.86 & 1.02 \\
         {[H]} & 0.75 & 0.75 & 0.18 & 0.18 \\
         {[He]} & 0.25 & 0.25 & $-$ & $-$\\
         {[C]} & $-$ & $-$ & 0.25 & 0.25 \\
         {[N]} & $-$ & $-$ & 0.27 & 0.27 \\
         {[O]} & $-$ & $-$ & 0.30 & 0.30\\
         {[${}^2$H]} & 7.5$\times 10^{-6}$ & 7.5$\times 10^{-6}$ & 1.8$\times 10^{-6}$ & 1.8$\times 10^{-6}$\\
         {[${}^3$He]} & 2.5$\times 10^{-5}$ & 2.5$\times 10^{-5}$ & $-$ & $-$ \\
         {[${}^{13}$C]} & $-$ & $-$ & 2.5$\times 10^{-5}$ & 2.5$\times 10^{-5}$ \\
         {[${}^{15}$N]} & $-$ & $-$ & 1.1$\times 10^{-3}$ & 1.1$\times 10^{-3}$ \\
         {[${}^{17}$O]} & $-$ & $-$ & 1.2$\times 10^{-4}$ & 1.2$\times 10^{-4}$\\ \bottomrule 
    \end{tabular}
    \label{tab:planet_data}
\end{table}

\begin{table}[h!]
    \centering
    \caption{Spin {$J$} and expectation values {$\ev{S_p}$ and $\ev{S_n}$} of the proton and neutron spins, for elements relevant for spin-dependent scattering.}
    \begin{tabular}{l c c c c c c c}
        \toprule 
        & H & ${}^2$H & ${}^3$He & ${}^{13}$C & ${}^{14}$N & ${}^{15}$N & ${}^{17}$O\\ \midrule
        $J$ & 1/2 & 1 & 1/2 & 1/2 & 1 & 1/2 & 5/2 \\
        $\ev*{S_p}$ & 0.5 & 0.5 & $-0.081$ & 0 & 0.5 & $-0.136$ & $-0.008$\\
        $\ev*{S_n}$ & 0 & 0.5 & 0.552 & $-0.183$ & 0.5 & 0.028 & 0.48\\\bottomrule 
    \end{tabular}
    \label{tab:isotopes}
\end{table}

{To assess the observational prospects for our scenario, we briefly discuss the detectability of UV airglow from local Super-Jupiters. Measuring airglow in the $\sim 700$–$1800~\text{\AA}$ band requires space-based UV spectroscopy. In practice, the most robust detections are expected for (i) a widely separated planet, where stellar UV contamination is reduced, or (ii) a free-floating planetary-mass object. For unresolved remote observations, the detectability follows the inverse-square scaling of the received flux. For a band-integrated airglow luminosity $L_{\rm UV}$, the observed flux at distance $d$ is
\begin{equation}
    F_{\rm UV} = \frac{L_{\rm UV}}{4\pi d^2} \ ,
\end{equation}
which implies an approximate detection horizon
\begin{equation}
    d_{\max}\simeq\left(\frac{L_{\rm UV}}{4\pi F_{\rm lim}}\right)^{1/2} \ ,
\end{equation}
where $F_{\rm lim}$ is the line or band flux sensitivity for the relevant exposure time and background. The numerical value of $d_{\max}$ is therefore instrument- and exposure-dependent; however, the accessible volume improves rapidly with UV sensitivity and collecting area.}

{Existing and archival observations with the Hubble Space Telescope and potentially GALEX can probe the nearest candidates, while future large-aperture UV mission concepts (e.g., LUVOIR/LUMOS- or HabEx-like capabilities) would substantially extend the reachable distance. Nearby planetary-mass objects have already been observed, such as WISE~J085510.83$-$071442.5 at $\sim 2.2$~pc, demonstrating that the Solar neighborhood contains cold substellar bodies. More broadly, local-volume censuses include numerous brown dwarfs and super-Jupiter–mass objects within $\sim 10$~pc, with ongoing wide-field surveys continuing to improve completeness. We therefore regard local Super-Jupiters, especially free-floating or wide-orbit systems, as plausible future targets for future UV airglow searches, and improved UV instrumentation would make such observations increasingly realistic.}

\section{Additional Cross Section Details and Results}

In this section, we explore various types of scattering interactions between DM and the elements composing the giant planets. The main types of interactions are spin-dependent, either through a proton or neutron scattering, and spin-independent. We address these separately and present the corresponding limits which can be obtained through UV airglow; note that the spin-dependent proton scattering limits are shown in Fig.~\ref{fig:H2_LW} of the main text.

\subsection{Spin-Dependent Scattering Cross Section}

For a nucleus with spin $J$, the conversion of the nucleon to nucleus scattering cross section is given by
\begin{align}\label{eq:SD_cross_section}
    \sigma_{\chi N}^\text{SD} =  \frac{\mu_{\chi N}^2}{\mu_{\chi j}^2} \frac{4(J+1)}{3J} \left[a_p\left\langle S_p \right\rangle + a_n \left\langle S_n \right\rangle \right]^2  \sigma_{\chi \lbrace p,n \rbrace}^\text{SD} \ ,
\end{align}
where ${\mu_{\chi N}}$ and ${\mu_{\chi j}}$ is the DM-nucleus and DM-nucleon reduced mass respectively, ${\left\langle S_p\right\rangle}$ and ${\left\langle S_n\right\rangle}$ are the average proton and neutron spin respectively, $a_p$ and $a_n$ are DM-proton and DM-neutron couplings respectively. When investigating DM-proton limits we assume $a_p=1$ and $a_n=0$, and for DM-neutron limits we assume $a_p=0$ and $a_n=1$.

\begin{figure}[htb]
    \centering
        \includegraphics[width=0.49\textwidth]{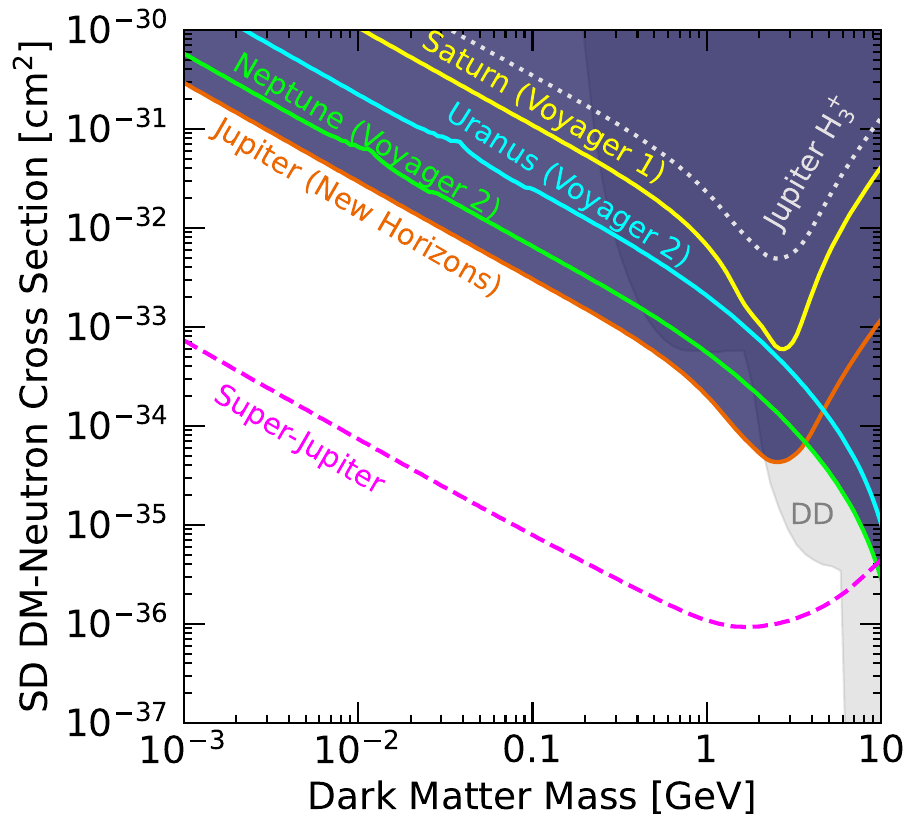}
        \includegraphics[width=0.49\textwidth]{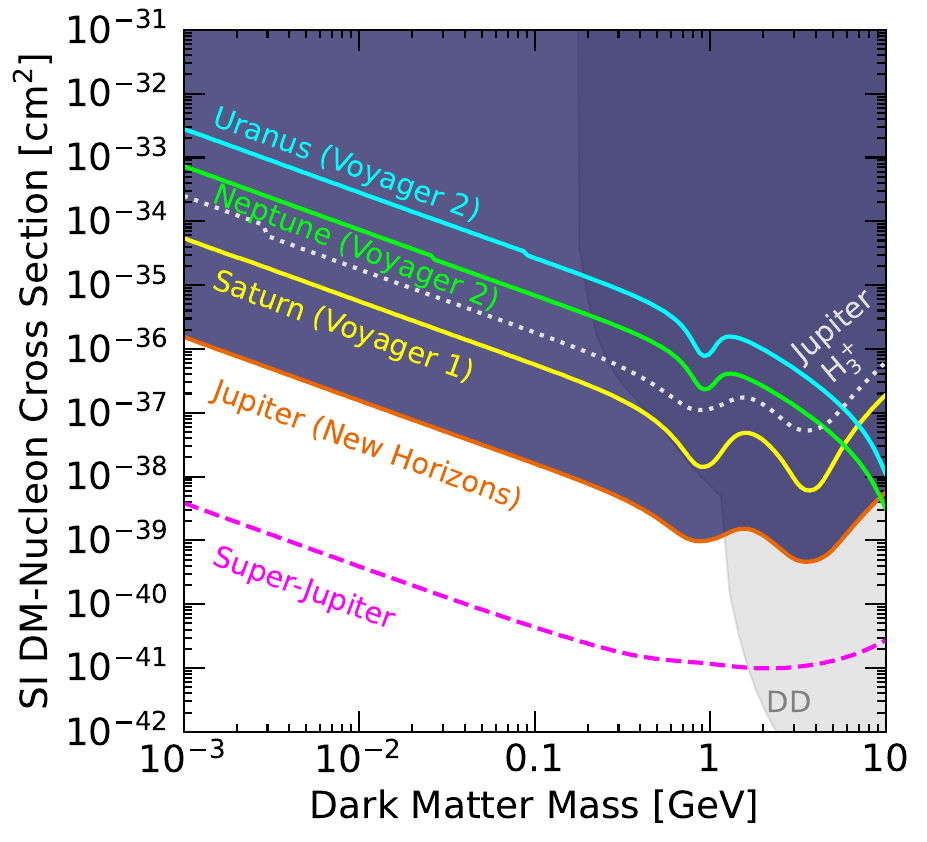}
    \caption{\label{fig:spin_dependent} New constraints on the DM-neutron spin-dependent {(left) and DM-nucleon spin-independent (right)} elastic scattering cross section, using DM-induced UV airglow in Jupiter, Saturn, Uranus, and Neptune. We also show projected sensitivity for a nearby Super-Jupiter. Complementary constraints from \hp production in Jupiter and direct detection (DD) experiments are shown as labeled.
    }
\end{figure}

{The left panel of Fig.}~\ref{fig:spin_dependent} shows our SD DM-neutron scattering limits for the gas and ice giants, as well as a projection for a benchmark (10 Jupiter mass and 1 Jupiter radius) local Super-Jupiter (see main text for more details on the benchmark). We incorporate direct detection constraints from CDMSlite~\cite{SuperCDMS:2018gro}, CRESST-III~\cite{CRESST:2022dtl}, and XENONnT~\cite{XENON:2023cxc}. The H$_3^+$ constraints are recast from Ref.~\cite{Blanco:2023qgi}. In this figure, no evaporation is included due to its model dependence~\cite{Acevedo:2023owd}, and all the DM annihilation products are assumed to be deposited into the planetary atmospheres; we investigate model-specific results in a companion paper~\cite{Blanco:2025wpo}.

For DM-proton scattering constraints, see Fig.~\ref{fig:H2_LW} of the main text. The location and depth of the sensitivity trough differs between proton and neutron spin-dependent cross sections; proton scattering primarily involves hydrogen and nitrogen, while neutron scattering involves ${}^3$He and nitrogen.

\subsection{Spin-Independent Scattering Cross Section}

We convert the spin-independent scattering of DM with planetary constituents from per-nucleon to per-nucleus cross section via
\begin{align}\label{eq:SI_cross_section}  
    \sigma_{\chi N}^\text{SI} =  \frac{\mu_{\chi N}^2}{\mu_{\chi j}^2} A^2 \sigma_{\chi j}^\text{SI} \ , 
\end{align}
where ${\sigma_{\chi j}^\text{SI}}$ is the DM-nucleon cross section, ${\mu_{\chi N}}$ and ${\mu_{\chi j}}$ is the DM-nuclei and DM-nucleon reduced mass respectively, and ${A}$ is the number of nucleons in {the} atom.

{The right panel of Fig.}~\ref{fig:spin_dependent} shows our new constraints on the spin-independent scattering cross section using DM-induced UV airglow. The direct detection scattering constraints are from CRESST-III~\cite{CRESST:2019jnq}, PandaX~\cite{PandaX:2022aac}, and DarkSide-50~\cite{DarkSide-50:2022qzh}. The H$_3^+$ constraints are recast from Ref.~\cite{Blanco:2023qgi}. There are also constraints on part of this parameter space from the early Universe as well as rare-Kaon decays, which depend on the particle physics model~\cite{Cox:2024rew}; see also Ref.~\cite{Knapen:2017xzo} for some other complementary constraints. In this figure, no evaporation is included due to its model dependence~\cite{Acevedo:2023owd}, and all the DM annihilation products are assumed to be deposited into the planetary atmospheres; we investigate model-specific results in a companion paper~\cite{Blanco:2025wpo}.

\newpage
\bibliography{main}
\bibliographystyle{apsrev4-2}

\end{document}